\documentclass{aa}
\usepackage{graphicx}

\begin{document}

\thesaurus{}

\title{Microlensing neutron stars}

\author{Dominik J. Schwarz \inst{1,2}
\and    Dirk Seidel \inst{2}}

\offprints{Dominik J. Schwarz}

\institute{Institut f\"ur Theoretische Physik,
           Technische Universit\"at Wien, Wiedner Hauptstra{\ss}e 8--10,
           A-1040 Wien \\
           email: dschwarz@hep.itp.tuwien.ac.at
\and       Institut f\"ur Theoretische Physik,
           J.W. Goethe-Universit\"at, D-60054 Frankfurt am Main \\
           email: seidel@th.physik.uni-frankfurt.de}

\date{Received 27 March 2001 / Accepted 28 March 2002}

\maketitle

\begin{abstract}
We investigate the chances that neutron stars act as the lens in a 
gravitational microlensing event towards the galactic bulge or a spiral arm. 
The observation of neutron stars by means of gravitational microlensing 
would allow the estimation of neutron star masses independently of the 
property of being a pulsar in a binary system. We estimate the contribution 
of neutron stars to the optical depth and the lensing rate based on two 
different models of pulsar distribution in the galaxy. Since only a small 
fraction of all neutron stars are pulsars, it is unlikely to find a pulsar 
that acts as a microlens by chance. A position comparison of known radio 
pulsars with observed microlensing candidates towards the galactic bulge and 
spiral arms shows no candidate pairs, which is consistent with the theoretical 
expectation. To improve the probability of microlensing a pulsar, we suggest a 
search for gravitational microlensing events of known nearby high proper 
motion pulsars. The pulsar \object{PSR J1932+1059} is a good candidate for
an astrometric detection of gravitational lensing.

\end{abstract}

\section{Introduction}

The determination of neutron star masses so far has been limited to
binary systems. Particularly, the masses of radio pulsars in binary systems
lie in an impressively narrow range $M_{\rm PSR} = 1.35 \pm 0.04 M_{\sun}$
(Thorsett et al.~\cite{T93}; Thorsett \& Chakrabarty \cite{T99}). The
spread of masses based on the measurement of X-ray binary systems is
somewhat larger (Nagase \cite{N89}; Reynolds, Bell \& Hilditch \cite{R92};
Heap \& Corcoran \cite{H92}; Reynolds et al.~\cite{R97}; 
Ash et al.~\cite{A99}), but the measurements are consistent with the mass value 
quoted above. However, it is not clear whether these measurements are 
representative of all neutron stars. In this paper we investigate the 
chances of estimating the mass of neutron stars by gravitational microlensing. 

The galaxy hosts $\sim 10^9$ neutron stars (Timmes, Woosley \& Weaver 
\cite{TWW96}), of which $\sim 10^5$ are pulsars (Narayan \cite{Narayan}). 
Thus the fraction of neutron stars which are also pulsars is only $10^{-3} - 
10^{-4}$. Moreover, only a small fraction of pulsars appears to be in binary 
systems (of $\sim 10^3$ known radio pulsars, only $\sim 10^2$ are 
known to be in binary systems). The range of observed masses is 
surprisingly narrow in view of the theoretical stability considerations, 
which allow neutron star 
masses in the range $0.1 M_{\sun}$ to $3 M_{\sun}$ (see, e.g., Weber 
\cite{Weber}). The predicted initial mass function from supernova 
simulations provides a more narrow mass range within $1 M_{\sun}$ to 
$2 M_{\sun}$, depending on the supernova type (Timmes, Woosley \& Weaver 
\cite{TWW96}).

The neutron star mass function is of great importance not only for
astrophysical purposes, but also for fundamental physics. On the one hand, 
neutron stars are the only known place in nature where theoretical ideas 
on the
phase diagram of quantum chromodynamics at high density may be tested
[for recent reviews see Schwarz (\cite{S98}), Weber (\cite{Weber}) and
Alford, Bowers \& Rajagopal (\cite{ABR00})]. On the other hand, the 
equation of state of neutron stars is important to predict the 
wave pattern of gravitational radiation emitted in mergers of neutron 
star binaries, one of the most important sources expected for gravitational 
wave interferometers like LIGO\footnote{ 
Laser Interferometer Gravitational Wave Observatory 
{\tt http://www.ligo.caltech.edu/}} and 
VIRGO\footnote{\tt http://www.virgo.infn.it/}.

Gravitational lensing [see, e.g., Schneider, Ehlers \& Falco 
(\cite{Schneider})] by now has been proven to be a very powerful 
technique to study mass distributions. For dark or low-luminosity stellar 
mass objects, gravitational microlensing [for reviews see Paczy\'nski 
(\cite{Paczynski}) and Zakharov \& Sazhin (\cite{ZS98})] provides a tool 
to measure their masses. Gravitational microlensing towards the 
galactic bulge and towards the galactic spiral arms has been studied by 
various groups (OGLE\footnote{The Optical Gravitational Lens Experiment
{\tt http:// astro.princeton.edu/\~{}ogle/}},
Udalski et al.~\cite{Udalski1}, \cite{Udalski2}, 
Wozniak et al.~\cite{Wozniak}; 
MACHO\footnote{\tt http://wwwmacho.mcmaster.ca/}, 
Alcock et al.~\cite{Alcock1}, \cite{Alcock2a}, \cite{Alcock2}, 
Popowski et al.~\cite{Popowski}; 
EROS\footnote{\tt http://eros.in2p3.fr/}
Derue et al.~\cite{Derue1}, \cite{Derue2}). The optical depth of the bulge
is $\tau = (1.4 \pm 0.3) \times 10^{-6}$ (Popowski et al.~\cite{Popowski})
and drops to $\tau = 0.45^{+0.24}_{-0.11} \times 10^{-6}$ 
(Derue et al.~\cite{Derue2}) in the spiral arms. About $800$ candidates for 
gravitational microlensing have been observed. 

In this work we show that gravitational microlensing of neutron stars in 
the galaxy can in principle be observed with present-day techniques
(photometric lensing). We find that the rate of observing a neutron 
star microlensing event is about $10^{-7}$ per year per observed light 
source in the direction of the galactic bulge; the optical depth is 
$\sim 10^{-8}$. Correspondingly, if we assume that the pulsar number 
density is proportional to the neutron star number density, the rate of 
observing a pulsar as a microlens is lower by a factor $10^{-3} - 10^{-4}$. 
As the number of monitored stars is $\sim 10^7$ in each experiment, we 
expect that no such observation has been made yet. A data 
comparison between the ATNF Pulsar Catalogue\footnote{\tt
http://wwwatnf.atnf.csiro.au/research/pulsar/ catalogue} and known microlensing 
events towards the galactic bulge and towards the galactic spiral arms 
(OGLE, Wozniak et al.~\cite{Wozniak}; MACHO, 
Alcock et al.~\cite{Alcock1}, \cite{Alcock2a}, \cite{Alcock2}; 
EROS, Derue et al.~\cite{Derue1}, \cite{Derue2}) shows 
that indeed none of the known pulsars has been acting as an observed 
microlens so far.

Once a microlensing event of appropriate duration (a few days to a 
few months) has been found the challenge is to figure out whether a 
neutron star or a different compact object has been acting as lens. 
This task may be accomplished by means of astrometric microlensing
(H{\o}g, Novikov \& Polnarev \cite{HNG}, Miyamoto \& Yoshi \cite{MY},
Walker \cite{W}, Miralda-Escud\'e \cite{M} and Boden, Shao \& Van Buren 
\cite{BSB}). Astrometric lensing measures the shift of the 
centre of light of the source star. While photometric lensing is most efficient
if the source passes through the Einstein ring, astrometric lensing is most 
efficient if the source passes close to, but outside, the Einstein ring. 
The most important background will be main sequence stars that are not 
properly resolved (Gould \cite{G00}). White or brown dwarfs as well as 
black holes are candidates for similar microlensing events due to their 
similar masses. As has been shown by Gould (\cite{G00}), upcoming astrometric 
space missions (like 
GAIA\footnote{\tt http://astro.estec.esa.nl/GAIA/}
and
SIM\footnote{Space Interferometry Mission {\tt http://sim.jpl.nasa.gov/}})
combined with ground-based observations should be able to 
pick out a handful of neutron stars. The observation of astrometric
lensing will also be possible with the help of long-baseline optical
interferometry, and has been studied in some detail for the ESO Very Large 
Telescope Interferometer 
(VLTI\footnote{\tt http://www.eso.org/projects/vlti/}) 
by Delplancke et al.~(\cite{DGR}). 
The ESO VLTI offers yet another new possibility; it might be possible for the 
first time to resolve the double images of microlensed stars. For events 
that are long enough to observe the parallax effect of the light-curve, a
measurement of the lens mass will be possible (Delplancke et al.~\cite{DGR}). 

In order to improve the lensing probability for neutron stars we propose 
a new search strategy that looks for the microlensing event of a 
known nearby high proper motion pulsar (sweeping a large angle on the 
sky). In that case the 
distance and velocity are constrained from the dispersion measure or via 
parallax measurement and a measurement of the pulsar mass should be 
possible. Thus gravitational microlensing might enable us to determine 
the mass of stand-alone pulsars. The cross section of astrometric
lensing is larger than that of photometric lensing, which 
additionally enhances the probability of detection. Thus astrometric 
measurements around high velocity proper motion pulsars in dense fields might 
allow a mass determination of pulsars. The potential of forthcoming astrometric
space missions for astrometric microlensing of ordinary stars has been 
investigated in detail by Paczy\'nski (\cite{P98}), Gould (\cite{G00}) 
and Salim \& Gould (\cite{SG}) for SIM and by Belokurov \& Evans (\cite{BE})
for GAIA. 

In section \ref{lrod} we use two pulsar population models of our galaxy
to estimate the lensing rate and the optical depth towards different
directions. In section \ref{comparison} we describe the catalogue 
comparison of pulsars and gravitational microlensing events. 
We suggest a search strategy to increase the probability to find a 
neutron star that acts as a gravitational microlense in section 
\ref{search} and end with a short conclusion. 

\section{Lensing rate and optical depth}\label{lrod}

Let us first investigate the possibility that a neutron star 
is detected by photometric microlensing.
 
\subsection{Lensing rate}\label{lr}

A simple estimate for the rate $p$ of gravitational microlensing events
per observed background star is obtained easily. We consider a neutron star
at distance $r$ that passes by the line of sight to a background star at
distance $D_*$ within the Einstein radius
\begin{equation}\label{R_E}
R_{\rm E}=\sqrt{\frac{2R_{\rm S}}{D_*}r(D_*-r)} \ ,
\end{equation}
where $R_{\rm S}$ denotes the Schwarzschild radius. The lensing rate is
proportional to the solid angle that is swept by the lens, which depends
on the (projected) velocity $v$ and the Einstein radius $R_{\rm E}$ of the
lens. Moreover, the lensing rate is proportional to the distance
$D_*$ of the source star and to the number density $n$ of lenses (neutron
stars), which we take as constant for this simple estimate. Let us for the 
moment assume that all neutron stars are distributed homogeneously in a 
cylinder with radius $10$ kpc and height $1$ kpc and that the number of
neutron stars in our Galaxy is $N_{\rm NS} = 10^9$. By putting 
$D_* = 10$ kpc, $r=D_*/2$, $M_{\rm NS} = 1 M_{\sun}$ and 
$v=200\,{\rm km}/{\rm s}$ we find
\begin{equation}\label{lrsimple}
p\sim D_* R_{\rm E} n v \sim 10^{-7}\, {\rm yr}^{-1}.
\end{equation}
Presently, a few tens of millions of stars have been monitored over some 
years by various collaborations. This suggests that about a handful 
of by now observed lenses should be neutron stars. This rough estimate
is consistent with the estimate given in Gould (\cite{G00}). The difficulty 
is to tell which of the observed microlensing events was due to 
a neutron star. There are only about $10^5$ galactic pulsars (Narayan 
\cite{Narayan}), which gives a pulsar lensing rate of the order 
$10^{-11}\, {\rm yr}^{-1}$. Typically, a present-day experiment will therefore 
not find microlensing events that are due to pulsars.

In order to calculate the lensing rate and the optical depth more precisely 
we have to take into account that the number density of neutron stars is
not constant within our galaxy. There are several models for the pulsar 
number density $n_{\rm PSR}$ in our galaxy and we assume that the 
neutron star number density $n$ is proportional to it
\begin{equation}\label{n_ns_psr}
n(R,z) =\frac{N_{\rm NS}}{N_{\rm PSR}}n_{\rm PSR}(R,z) \ ,
\end{equation}
where $R$ is the radial distance of the pulsar/neutron star to the galactic
centre in the galactic plane and $z$ is the height. 

\begin{figure}[htb]
\resizebox{\hsize}{!}{\includegraphics{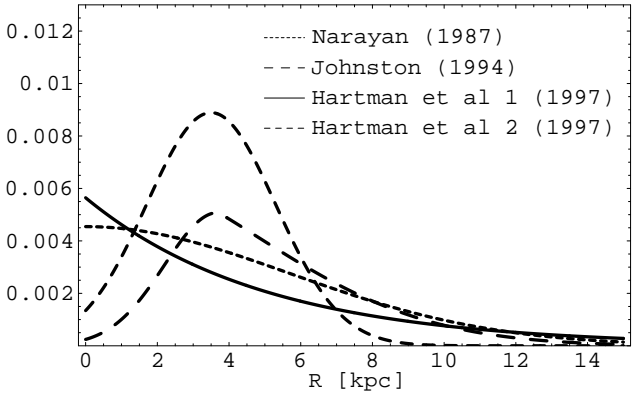}}
\caption{Four models for the radial distribution of neutron stars in the 
galactic disk. All four distributions are normalised to unity.}
\label{fourmodels}
\end{figure}

Following Hartman et al.~(\cite{Hartman}) we use two different models for the
radial dependence of the number density and a model for the 
$z$-dependence, which was given by Lyne et al.~(\cite{Lyne}). 
The two models from Hartman resemble the models from Narayan
(\cite{Narayan}) and Johnston (\cite{Johnston}). Figure
\ref{fourmodels} shows the number densities of the four models in the galactic
plane plotted against the distance from the galactic centre.
%%%%%%%%%%%%%%%%%%%%%%%%%%%%%%%%%%%%%%%%%%%%%%%%%%%%%%%%%%%%%%%%%%%%%%%%%%%
%%%%%%%%%%%%%%%%%%%%%%%%%%%%%%%%%%%%%%%%%%%%%%%%%%%%%%%%%%%%%%%%%%%%%%%%%%%

We use the assumption (Narayan \cite{Narayan}), that the number density can
be factorised
\begin{equation}
n(R,z) 2 \pi\, R {\rm d}R\, {\rm d}z=
N_{{\rm NS}}\left[n_R(R)\, 2 \pi\, R {\rm d}R\right]\left[n_z(z){\rm d}z\right].
\label{nfactor}
\end{equation}
The first model from Hartman is exponential in $R$,
\begin{equation}
n_{\rm H1}(R)=\frac{1}{2 \pi R_{\rm W}^2}\exp\left(-\frac{R}{R_{\rm W}}\right),
\label{nH1}
\end{equation}
where $R_{\rm W}=5\,{\rm kpc}$.
There is some evidence for a deficit of pulsars in the galactic centre 
which was shown by Johnston (\cite{Johnston}). The second model 
from Hartman takes this into account
\begin{equation}
n_{\rm H2}(R)=
\frac{c_{\rm H2}}{2\pi R_{\rm w}^2}
\exp\left(-\frac{\left(R-R_{{\rm max}}\right)^2}{2R_{\rm w}^2}\right),
\label{nH2}
\end{equation}
where $R_{\rm w}=1.8$ kpc and $R_{\rm max} = 3.5$ kpc.
The normalisation constant is $c_{\rm H2} \approx 0.204$ for the given choice
of $R_{\rm max}$. For Gaussian distribution around the centre $c_{\rm H2} = 1$.
For the $z$-dependence we use the Gaussian
\begin{equation}
n_z(z):=
\frac{1}{\sqrt{2\pi}\sigma}
\exp\left(-\frac{1}{2}\frac{z^2}{\sigma^2}\right)
\label{nL}
\end{equation}
with $\sigma=0.45\,{\rm kpc}$ (Lyne et al.~\cite{Lyne}).

We transform the equations ({\ref{nH1}), ({\ref{nH2}) and ({\ref{nL}) into
spherical coordinates $(r,\theta,\phi)$ with the sun at the origin via
\begin{eqnarray}
z &=& r\sin\theta, \label{ztrans} \\
R^2 &=& r^2\cos^2\theta+R_{{\rm SC}}^2-2rR_{{\rm SC}}\cos\theta\cos\phi.
\label{transR}
\end{eqnarray}
The distance from the sun to the centre of the galaxy is given by
$R_{{\rm SC}}=8.5\,{\rm kpc}$.

Hansen and Phinney (\cite{Hansen}) suggested a Maxwell distribution for
the kick velocities of neutron stars:
\begin{equation}
f(v)=
\sqrt{\frac{2}{\pi}}\frac{v^2}{\sigma_v^3}
\exp\left(-\frac{v^2}{2\sigma_v^2}\right)
\label{HaP}
\end{equation}
where $\sigma_v=190\,{\rm km s}^{-1}$, which corresponds to a mean velocity 
at birth of $\sim 300 \,{\rm km s}^{-1}$. For pulsars that are older than 
$10^7$ years they find a mean velocity of $198 \pm 53 \, {\rm km s}^{-1}$.
We can now calculate the lensing rate, which is given by
\begin{equation}
p(D_*,\theta,\phi) = N_{\rm NS} \pi v_{\perp} 
\int_0^{D_*} R_{\rm E}(r, D_*) n(r,\theta,\phi) {\rm d}r\ .
\label{lensingrate}
\end{equation}
It only depends on the part of the velocity vector that lies in the
lens plane. Therefore we have to calculate the mean velocity projection on
the lens plane $v_{\perp}$ which is given by
\begin{equation}
v_\perp = \frac{1}{\pi}
          \int_{-\frac{\pi}{2}}^{\frac{\pi}{2}} v \cos\alpha\, {\rm d}\alpha 
        = \frac{2v}{\pi}.
\label{vperp1}
\end{equation}
Figures \ref{lrphi} and \ref{lrD_s} show the lensing rate as a function of the
galactic longitude [$\phi \equiv (l/180\degr) \pi$] and as a function of the 
distance to the light source, respectively. It is most promising 
to search for neutron stars that act as microlense in the direction of the
galactic bulge, but it is also reasonable to look into the disk at galactic 
longitudes below $50\degr$. 
%%%%%%%%%%%%%%%%%%%%%%%%%%%%%%%%%%%%%%%%%%%%%%%%%%%%%%%%%%%%%%%%%%%%%
\begin{figure}[htb]
\resizebox{\hsize}{!}{\includegraphics{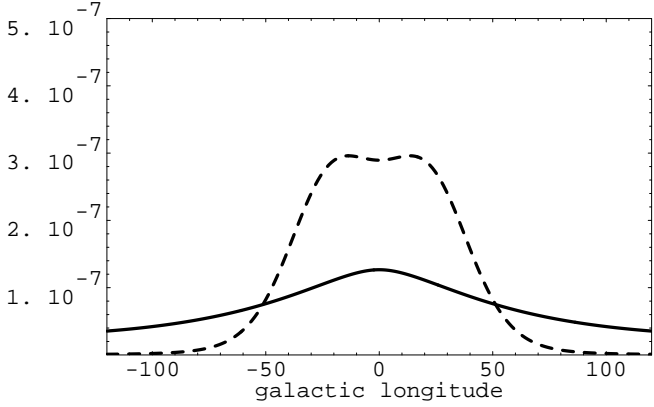}}
\caption{Lensing rate in units of number of events/source/year
for light sources in the galactic plane at a distance 
of $D_* = 7.5$ kpc, as a function of the galactic longitude. The two 
estimates are
based on two models for the radial distribution of pulsars by 
Hartman et al.~(\cite{Hartman}). The solid (dashed) line represents the 
model H1 (H2).} 
\label{lrphi}
\end{figure}
%%%%%%%%%%%%%%%%%%%%%%%%%%%%%%%%%%%%%%%%%%%%%%%%%%%%%%%%%%%%%%%%%%%%%%%
\begin{figure}[htb]
\resizebox{\hsize}{!}{\includegraphics{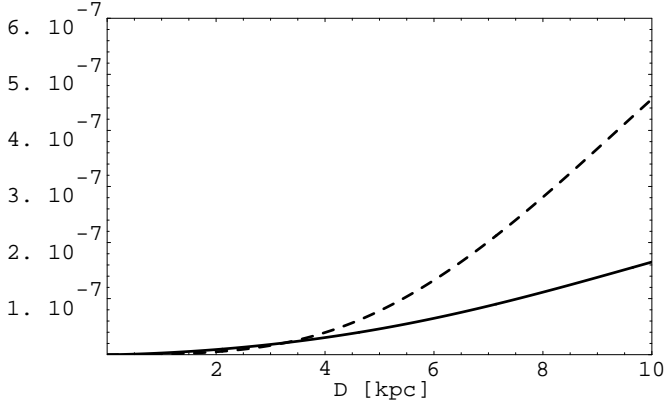}}
\caption{Lensing rate in units of number of events/source/year
for light sources at galactic longitude $30\degr$ 
in the galactic plane, as a function of distance $D_*$, for the same models
as in Fig. \ref{lrphi}.}
\label{lrD_s}
\end{figure}
%%%%%%%%%%%%%%%%%%%%%%%%%%%%%%%%%%%%%%%%%%%%%%%%%%%%%%%%%%%%%%%%%%%%%%%

\subsection{Optical depth}\label{od}

Analogous to (\ref{lrsimple}) we estimate the order of the optical depth, 
\begin{equation}
\label{odsimple}
\tau\sim n D_* R_{\rm E}^2 \sim\frac{p R_{\rm E}}{v}\sim p\, 0.1\, {\rm yr}
\sim 10^{-8}\ .
\end{equation}
%%%%%%%%%%%%%%%%%%%%%%%%%%%%%%%%%%%%%%%%%%%%%%%%%%%%%%%%%%%%%%%%%%%%%%%%
\begin{figure}[htb]
\resizebox{\hsize}{!}{\includegraphics{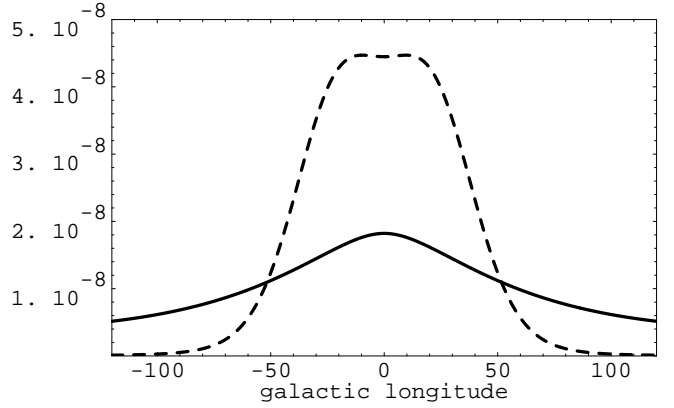}}
\caption{Optical depth for light sources in the galactic plane at a distance
of $D_* = 7.5$ kpc, as a function of the galactic longitude, for the same 
models as in Fig. \ref{lrphi}.}
\label{odphi}
\end{figure}
%%%%%%%%%%%%%%%%%%%%%%%%%%%%%%%%%%%%%%%%%%%%%%%%%%%%%%%%%%%%%%%%%%%%%%%%%
\begin{figure}[htb]
\resizebox{\hsize}{!}{\includegraphics{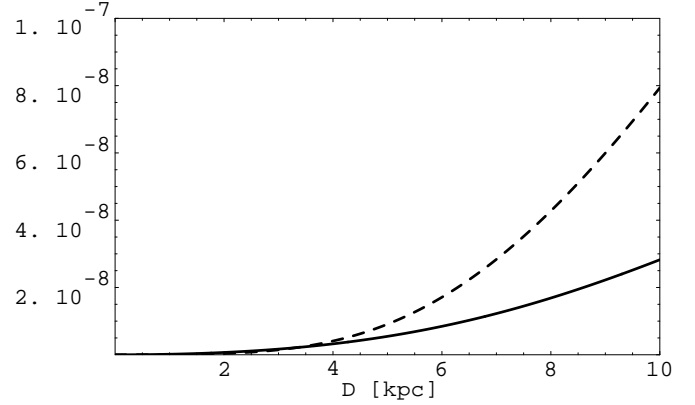}}
\caption{Optical depth for light sources at galactic longitude $30\degr$
in the galactic plane, as a function of distance $D_*$, for the same models
as in Fig. \ref{lrphi}.}
\label{odD_s}
\end{figure}
%%%%%%%%%%%%%%%%%%%%%%%%%%%%%%%%%%%%%%%%%%%%%%%%%%%%%%%%%%%%%%%%%%%%%%%%%%%
A more thorough estimate for the optical depth is obtained from 
\begin{equation}
\tau(D_*,\theta,\phi)=\int_0^{D_*}\pi R_{\rm E}^2(r,D_*) 
n(r,\theta,\phi){\rm d}r.
\label{opticaldepth}
\end{equation}
Figures \ref{odphi} and \ref{odD_s} show the optical depth as a function 
of the galactic longitude and as a function of the distance to the light 
source, respectively.

\subsection{Typical timescale}

Consider a lens of mass $M$, passing the line of sight of a distant 
source at a perpendicular velocity $v_\perp$. The typical time 
scale $t_{\rm var}$ of the microlensing event is given by 
\begin{eqnarray}
t_{\rm var} &\equiv& \frac{R_{\rm E}}{v_\perp} \nonumber \\ 
&\approx& 25\,{\rm d}\left(\frac{M}{M_{\sun}}\right)^{1/2}
\left(\frac{D}{1\,{\rm kpc}}\right)^{1/2}
\left(\frac{200\,{\rm km/s}}{v_\perp}\right),
\label{tvar}
\end{eqnarray}
where $D = r(D_* - r)/D_*$. For $M = 1.4 M_{\sun}$ we find that the  
duration of a neutron star microlensing event might range from a few days
for close and fast neutron stars up to a few months for neutron stars close
to the galactic bulge. 

The distribution of timescales is estimated in figure \ref{duration}.
We plot
\begin{eqnarray}
{{\rm d} p(t_{\rm var}) \over {\rm d} \ln t_{\rm var}} &=&
{2 N_{\rm NS} \over t_{\rm var}^2} \nonumber \\
& \times &  
\int_0^{D_*} R_{\rm E}^3(r,D_*) f\left[R_{\rm E}(r,D_*)\over t_{\rm var}\right] 
n(r,\theta,\phi) {\rm d} r
\end{eqnarray}
at $\theta = \phi = 0$ for $D_* = 3.25$ kpc and $D_* = 7.5$ kpc, respectively. 
We use the velocity distribution (\ref{HaP}), since we expect 
young neutron stars at small scale height.  
%%%%%%%%%%%%%%%%%%%%%%%%%%%%%%%%%%%%%%%%%%%%%%%%%%%%%%%%%%%%%%%%%%%%%
\begin{figure}[htb]
\resizebox{\hsize}{!}{\includegraphics{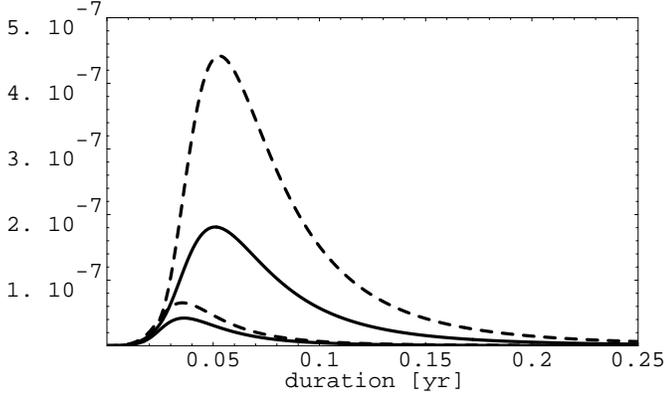}}
\caption{Timescale distribution towards the galactic centre for light 
sources at $D_* = 7.5$ kpc (upper lines) and $D_* = 3.25$ kpc (lower 
lines), for the same models as in Fig. \ref{lrphi}.
}
\label{duration}
\end{figure}
%%%%%%%%%%%%%%%%%%%%%%%%%%%%%%%%%%%%%%%%%%%%%%%%%%%%%%%%%%%%%%%%%%%%%%%
As can be seen from the comparison of the timescale distributions for sources
close to the galactic centre and sources at moderate distances, the
width of the timescale distribution is dominated by the sources close to the 
galactic centre. In both models for the radial distribution the timescale 
distribution peaks at durations of $\sim 20$ days.   

\subsection{Astrometric microlensing}

Astrometric microlensing makes use of the shift of the 
centroid of the combined (unresolved) images of the light source.
While the light amplification used in the photometric effect drops 
outside the Einstein ring as
\[
A \sim 1 + 2 \left(\theta_{\rm E}\over \theta_{\rm sl}\right)^2,
\]
the centroid shift is given by 
\[
\Delta \theta \sim {\theta_{\rm E}^2 \over \theta_{\rm sl}}.
\]
Here $\theta_{\rm E} \equiv R_{\rm E}/r$ and $\theta_{\rm sl}$ is the 
angle between the source and the lense. For a nearby neutron star we may assume 
$D_* \gg r$, thus $\theta_{\rm E} \approx 9 (M/M_{\sun})^{1/2} 
(100 \mbox{\ pc}/r)^{1/2}$ mas, giving rise to a typical centroid shift of
\begin{equation}
\label{astrometric}
\Delta \theta \approx 81
\left( M\over M_{\sun} \right) \left( 100 \mbox{\ pc}\over r\right) 
\left( 1\arcsec \over \theta_{\rm sl}\right) \mu\mbox{as}.
\end{equation} 
GAIA will achieve an astrometric accuracy of $11 \mu$as at $V = 15$ mag, 
which drops to $160 \mu$as at $V = 20$ mag (Perryman et al.~\cite{Pea}), 
while SIM will resolve $4 \mu$as for sources as faint as that 
magnitude (Boden, Unwin \& Shao \cite{BUS}). With the help of VLTI an 
astrometric accuracy of the order of $10 \mu$as can be expected 
(Delplancke, G\'orski \& Richichi \cite{DGR}) with a limiting magnitude $K = 
19 - 20$ (available from the middle of 2004). Thus it will be 
possible, at least in principle, to measure the centroid shift of 
background stars separated less than a few $\arcsec$ from a nearby neutron star.

For a survey mission like GAIA, the optical depth and timescale of lensing
may be estimated as above, simply correcting for the larger 
cross section due to the astrometric effect. If one adopts an accuracy of 
$\sim 100 \mu$as, and aims at measuring the centroid shift with an error of 
$10\%$, $\theta_{\rm sl}/\theta_{\rm E}$ might be as big as $10$, thus 
increasing the cross section from photometric lensing by a factor of $100$ 
and the time scale by a factor of $10$. However,
the sampling of GAIA will be sparse compared with ground-based ongoing
microlensing experiments and only a small fraction of all astrometric 
microlensing events will be detected (Belokurov \& Evans \cite{BE}). 

On the other hand SIM will focus on astrometry of preselected targets.
For photometrically-detected microlensing events it will be possible to 
detect the centroid shift and to obtain high accuracy mass determinations. 
 
\section{Catalogue comparison}\label{comparison}

%%%%%%%%%%%%%%%%%%%%%%%%%%%%%%%%%%%%%%%%%%%%%%%%%%%%%%%%%%%%%%%%%%%%%%
\begin{table}
\begin{center}
\begin{tabular}{|l|r|}
\hline
collaboration & number of microlensing events \\
\hline
OGLE  & bulge 520 \\
MACHO & bulge 144 \\
EROS  & bulge 19 \\
      & spiral arms 7 \\
      & total 690 \\
\hline
collaboration & number of radio pulsars \\
\hline
ATNF      & 1308 \\
\hline
\end{tabular}
\end{center}
\caption{\label{MLEventPulsar} Catalogue comparison of the coordinates of 
gravitational microlensing events and radio pulsars.}
\end{table}
%%%%%%%%%%%%%%%%%%%%%%%%%%%%%%%%%%%%%%%%%%%%%%%%%%%%%%%%%%%%%%%%%%%%%%%%%

As a consistency check of the above estimates of the photometric lensing 
rates we investigate whether any pulsar has acted as a microlens thus far.

The positions of 690 candidates for gravitational microlensing events 
towards the galactic bulge and the spiral arms (OGLE\footnote{
Catalogue of 520 Microlensing Events in the Galactic Bulge: \\
{\tt http://astro.princeton.edu/\~{}wozniak/dia/lens/}}, 
Wozniak et al.~\cite{Wozniak}; 
MACHO\footnote{\tt http://wwwmacho.mcmaster.ca/Data/bulgefts.html}, 
Alcock et al.~\cite{Alcock1}, \cite{Alcock2a}, \cite{Alcock2}; 
EROS\footnote{alerts 1998 - 2000,
{\tt http://www-dapnia.cea.fr/Phys/Spp/ Experiences/EROS/alertes.html}},
Derue et al.~\cite{Derue1}, \cite{Derue2})
have been compared with the positions of 
1308 radio pulsars from the ATNF Pulsar Catalogue\footnote{\tt 
http://wwwatnf.atnf.csiro.au/research/pulsar/ catalogue}, which is based on 
the Princeton Pulsar Catalogue\footnote{
\tt http://pulsar.princeton.edu/pulsar/catalog.shtml} (Taylor, Manchester \& 
Lyne \cite{Taylor}), 
the Parkes Multibeam Pulsar Survey\footnote{
\tt http://wwwatnf.atnf.csiro.au/research/pulsar/ pmsurv/}, 
(Manchester et al.~\cite{Manchester}) and the Swinburne Intermediate 
Latitude Pulsar Survey\footnote{\tt http://astronomy.swin.edu.au/pulsar/} 
(Edwards et al.~\cite{Edwards}). 
Table \ref{MLEventPulsar} lists the number of records of the individual groups.
Some of the microlensing events have been detected by more than one group.
We did not account for this in the table.
%%%%%%%%%%%%%%%%%%%%%%%%%%%%%%%%%%%%%%%%%%%%%%%%%%%%%%%%%%%%%%%%%%%%%%%%%%%%%%
\begin{table*}
\begin{center}
\begin{tabular}{|l|rrr|rrrrr|r|r|}
\hline
PSR J & 
$\mu_{\rm ra}$ & 
$\mu_{\rm dec}$ & 
$|\mu|$ & 
ra & 
dec & 
$l$ & 
$b$ & 
epoch &
$r$ & 
sources \\ 
 & mas/yr & mas/yr & mas/yr & h : m : s & $\degr$ : $\arcmin$ : $\arcsec$ & 
$\degr$ & $\degr$ & yr & kpc & $\theta_{\rm sl} < 15\arcsec$  \\
\hline
0437-4715 &  
$114 \pm 2$ & $-72 \pm 4$ & $106$ & 
04:37:15.71 & $-$47:15:07.0 & $253$ & $-42$ & 1993 & $0.14$ & 1/0 \\ 
0826+2637 & 
$61 \pm 3$ & $-90 \pm 2$ & $105$ & 
08:26:51.31 & $+$26:37:25.6 & $172$ & $35$ & 1969 & $1.25$ & 0/0 \\ 
1136+1551 & 
$-102 \pm 5$ & $357 \pm 3$ & $370$ & 
11:36:03.30 & $+$15:51:00.7 & $240$ & $69$ & 1975 & $0.27$ & 0/0 \\ 
1239+2453 & 
$-106 \pm 4$ & $42 \pm 3$ & $105$ &  
12:39:40.47 & $+$24:53:49.3 & $252$ &  $87$ & 1985 & $0.56$ & 0/0 \\ 
1932+1059 &  
$99 \pm 6$ & $39 \pm 4$ & $105$ & 
19:32:13.90 & $+$10:59:32.0 & $47$  &  $-4$ & 1991 & $0.17$ & 4/5 \\ 
2225+6535 &  
$144 \pm 3$ & $112 \pm 3$ & $127$ & 
22:25:52.36 & $+$65:35:33.8 & $109$ & $7$ & 1991 & $2.00$ & 1/1 \\
\hline
\end{tabular}
\end{center}
\caption{All pulsars with measured proper motion $|\mu| > 100$ mas/yr and 
$|\Delta \mu_{\rm ra,dec}| < 10$ mas/yr from the Princeton Pulsar 
Catalogue. Coordinates are given in the J2000 system at the specified epoch.
For catalogue entries without position epoch we quote the period epoch of 
the respective pulsar. The distance estimates are based on the dispersion 
measure. The number of objects within $15\arcsec$ of a pulsar is obtained with 
the help of VizieR from the USNO-A2.0/GSC2.2.1 catalogues.}  
\end{table*}
%%%%%%%%%%%%%%%%%%%%%%%%%%%%%%%%%%%%%%%%%%%%%%%%%%%%%%%%%%%%%%%%%%%%%%%%%%%%

All positions are given in the J2000 coordinate system. However, not all of
the measurements were made at the same time. Therefore we have to 
consider a change of the pulsar position that will lead to a deviation angle
$\alpha \approx vt/r$, where $t$ is the time between the pulsar detection
and the microlensing event, $v_\perp$ is the (projected) pulsar velocity 
and $r$ the pulsar distance. For a conservative estimate that includes 
pulsars that are close, fast, and have been detected long ago, we take
$r = 100$ pc, $v =500$ km/s, $t=10$ yrs, which gives
$\alpha\approx 10\arcsec$. We found no candidate pair with an angular separation
below this value. In a second step we looked for the closest pair,
which is PSR J1807-2715 and the OGLE microlensing event sc18 2757. 
The angular separation is $0.08\deg$. According to the Priceton Pulsar 
Catalogue PSR J1807-2715 is at an estimated distance of $9.67$ kpc and no
proper motion is reported. Thus it is highly unlikely that it can be 
responsible for the OGLE event sc18 2757. 
These findings are in agreement with our estimate, which predicts an 
extremely small 
lensing rate for pulsars, i.e. $\sim 10^{-4}$/yr for $10^7$ monitored stars.

\section{A new search strategy}\label{search}

We have shown that probably some of the gravitational 
microlensing events towards the galactic bulge are due to neutron stars. 
For a specific microlensing event with a duration of a few days to a few 
months, one is interested in finding out whether the lens is a neutron 
star or a different object. There is not very much hope that with present 
day experiments a microlensing event will be detected whose lens is a 
known pulsar. Our catalogue comparison has confirmed this result. 
To raise the low probability it is clear that a better search strategy is 
needed.
With the prospects of astrometric lensing, we can expect that high 
precision measurements of masses will become possible, but still the problem
of identifying the nature of the lense will remain, especially for masses 
of $1 - 3 M_{\sun}$, since many possible objects are stable in that mass range.

%%%%%%%%%%%%%%%%%%%%%%%%%%%%%%%%%%%%%%%%%%%%%%%%%%%%%%%%%%%%%%%%%%%%%%%%%%%%
\begin{table*}
\begin{center}
\begin{tabular}{|l|l|l|l|l|l|l|}
\hline 
USNO-A2.0 & x[$\arcsec$] & y[$\arcsec$] & blue[mag] & red[mag] & epoch[yr] 
& id \\
\hline
0975-15208581 & $-$1.58 & $-$3.91 & 15.0 & 13.2 & 1954 & 1 \\
0975-15209296 &    4.68 &    5.04 & 19.3 & 17.7 & 1954 & 2 \\
0975-15208295 & $-$3.96 &    7.60 & 20.2 & 19.6 & 1954 & 3 \\
0975-15207539 & $-$10.23 &   6.01 & 20.2 & 19.5 & 1954 & 4 \\ 
\hline
GSC2.2.1 & x[$\arcsec$] & y[$\arcsec$] & blue[mag] & red[mag] & epoch[yr] 
& id \\
\hline
N0232103208857 & $-$2.33 & $-$3.06 & & 14.22$\pm$0.40 & 1993 & 5 (1?) \\
N0232103780    & $-$1.93 & $-$4.66 & 14.85$\pm$0.42 & 13.44$\pm$0.40 & 1993 
& 6 (1?) \\
N023210364008  &    4.66 &    4.86 & & 17.45$\pm$0.44 & 1992 & 7 (2?) \\
N0232103208858 &    4.82 &    5.08 & & 18.10$\pm$0.41 & 1993 & 8 (2?) \\
N023210363839  & $-$11.30 & $-$5.29 & 18.68$\pm$0.42 & 17.11$\pm$0.44 & 1992 
& 9 \\
\hline 
\end{tabular}
\end{center}
\caption{Objects from the USNO-A2.0 and GSC2.2.1 catalogues within 
$15\arcsec$ of PSR J1932+1059 at position epoch 1991. $x$ and $y$ denote the
relative positions to the pulsar in the arc projection. The astrometric 
(photometric) 
accuracy of the USNO-A catalogue is typically $0.25\arcsec (0.25 {\rm\ mag})$. 
The astrometric accuracy of GSC2.2 is typically $0.3\arcsec$, the photometric 
errors are included in the table. We refer to the objects in the text 
by the number in the last column of the table. Objects 1 and 2 seem both
to split into two objects (5 \& 6 and 7 \& 8 resp.) in GSC2.2. Objects 
3 and 4 are both beyond the magnitude limit of GSC2.2, object 9 does not 
have a counterpart in USNO-A2. Thus we find 7 objects in total.} 
\end{table*}
%%%%%%%%%%%%%%%%%%%%%%%%%%%%%%%%%%%%%%%%%%%%%%%%%%%%%%%%%%%%%%%%%%%%%%%%%%%%%

A unique identifier of a neutron star is the possibility that it is 
observed as a radio pulsar. 
Instead of choosing a field of background stars we suggest to choose a
pulsar and to monitor the background stars around it. The solid
angle $\Omega$ that is swept during $\Delta t$ by the Einstein-ring of a 
moving Schwarzschild-lens is given by 
\begin{equation}
\Omega= \frac{1}{4\pi r^2} 
\left(\pi R_{\rm E}^2 + 2 R_{\rm E} v_\perp \Delta t\right) \simeq 
\frac{R_{\rm S}}{2r}
\left(1 + \frac{2\Delta t}{\pi t_{\rm var}}\right) ,
\label{Omega}
\end{equation}
for $r\ll D_*$. Our strategy therefore must be to observe distant 
background stars in the direction of a close, fast-moving neutron star. 
Table 2 shows some high proper motion pulsars  
from the Princeton Pulsar Catalogue. We include all pulsars whose proper 
motion have been measured to be $|\mu| > 100$ mas/yr and keep only those with 
proper motion estimates better than $10\%$. We expect that pulsars at low 
galactic latitude have the highest probablility of acting as a gravitational 
lens. This expectation is confirmed by the result of a search for appropriate 
background stars, based on the USNO-A2.0 (Monet et al.~\cite{Monet}) and 
GSC2.2.1\footnote{\tt http://www-gsss.stsci.edu/gsc/gsc2/GSC2home.htm} 
catalogues. These catalogues contain $526,280,881$ resp.~$455,851,237$ sources 
and both cover the entire sky. The USNO-A2.0 catalogue above declination 
$-30\degr$ (only those plates are relevant in our context, see below) goes 
as deep as $21$ in $O$ (blue plates) and $20$ in $E$ (red plates), whereas the 
GSC2.2.1 catalogue is limited by $F=18.5$ in the red and $J=19.5$ in the blue. 
We search in a radius of $15\arcsec$. The number 
of hits are shown in the last column of table 2 for USNO-A2.0/GSC2.2.1.
This procedure selects three pulsars as interesting candidates for lensing.

The only candidate at high galactic latitude, \object{PSR J0437-4715}, is a 
binary pulsar and the USNO-A2.0 object 0375-01566415 (epoch 1981) is 
separated by only $\approx 0.5\arcsec$ after the measured pulsar proper 
motion is 
used to extrapolate to epoch 1981. Thus the USNO object seems to be identical
with the binary companion identified by Bell, Bailes \& Bessell \cite{BBB}, 
probably a white dwarf with $R = 20.1$, since they find no other star brighter 
than $R=23.8$ within $6\arcsec$ of the radio position. 

The closest object (USNO-A2.0 object 1500-09063417, GSC2.2.1 object 
N013123248640) to \object{PSR J2225+6535} is separated by $14.7\arcsec$ and 
thus is too far away from the pulsar for significant lensing.

The most promising candidate is the pulsar \object{PSR J1932+1059} (B1929+10).
This pulsar is at a (dispersion measure) distance of about $0.17\,{\rm kpc}$ 
(Princeton Pulsar Catalogue) and moves $105$ mas/yr. It lies towards 
the galactic disk ($l = 47.4\degr, b = -3.9\degr$) and is found to have 
$7$ objects brighter than red magnitude $\sim 20$ within $15\arcsec$. From the 
$9$ hits shown in table 3, $2$ objects are counted twice (see caption of 
table). In figure \ref{figps} (top panel) we map $9\arcsec$ around the pulsars
radio position in 1991. The crosses indicate its estimated position  
in the years 2001, 2011 and 2021. The pulsar moves away from the objects 
1,5,6 and towards 2,7,8 and remains at an almost constant distance to 3. At 
epoch 2001(2011,2021) objects 5 and 7 are separated from the pulsar by 
$4.8\arcsec (5.7\arcsec, 6.8\arcsec)$ and $5.9\arcsec (4.9\arcsec, 
4.0\arcsec)$, respectively.

In 1994, Pavlov et al.~(\cite{Pavlov}) identified the 
optical counterpart of PSR J1932+1059 using the Hubble Space Telescope. 
The offset from the estimated radio position at epoch 1994.5 was $0.39\arcsec$.
As a by-product they found 6 additional stars within $4\arcsec$ of the radio 
position. We show their positions (from table 5 of Pavlov et al.~\cite{Pavlov})
in the bottom panel of figure 7, which is a magnification of the top panel by 
a factor of 2. The pulsar is almost in the middle between HST objects 1 and 2 
and approaches HST object 5. 
At 2010 (2015) the pulsar will be separated from HST objects 1, 2 and 5 by 
$2.66\arcsec (3.16\arcsec)$, $1.74\arcsec(1.95\arcsec)$ and $1.81\arcsec
(1.35\arcsec)$, respectively.
According to Pavlov et al.~(\cite{Pavlov}), objects 1--3, 5 and 
6 are perhaps very distant white dwarfs strongly reddened by interstellar 
extinction. 

The angular scale of the Einstein radius of PSR J1932+1059 is $8.2$ mas, thus 
there is no photometric lensing effect. However, there is a chance to detect 
the astrometric effect. Let us first consider the possibility of detecting 
lensing with the help of the objects from table 3. These objects are 
considerably 
brighter than the six HST objects and therefore astrometric measurements 
should be simpler. According to equation 
(\ref{astrometric}) the images of 5 and 7 move between 2010 and 2020 
by $2 \mu$as and $3 \mu$as. The pulsar and the objects 5 and 7 are to a good 
approximation on a straight line; the centroid shift is directed 
along the same line. The accuracy of SIM will be about $4 \mu$as, thus 
the effect seems to be just below what can be detected. 

%%%%%%%%%%%%%%%%%%%%%%%%%%%%%%%%%%%%%%%%%%%%%%%%%%%%%%%%%%%%%%%%%%%%%%%%%%%%
\begin{figure}[htb]
\begin{center}
\includegraphics[width=0.8\linewidth]{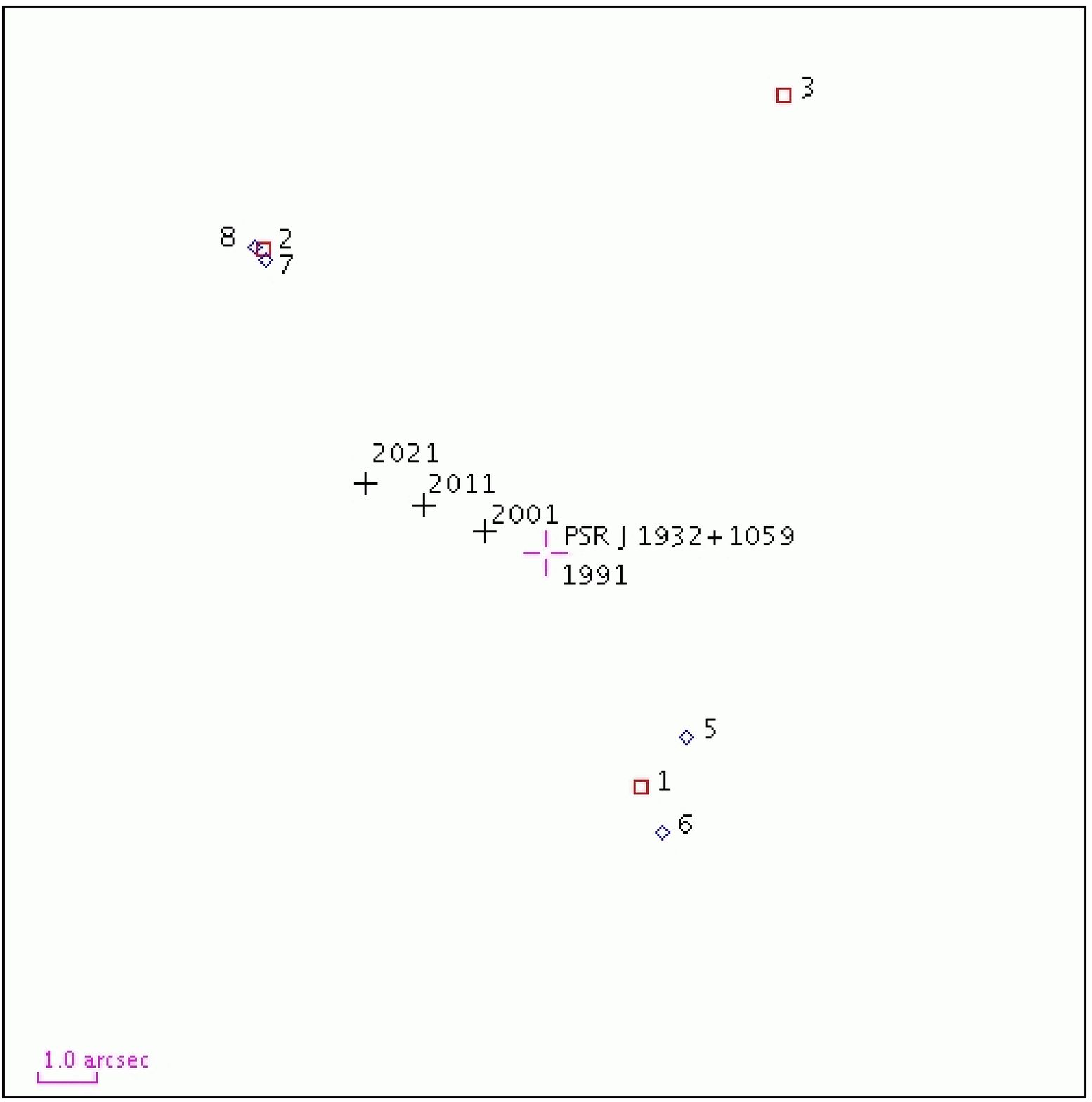}\\
\includegraphics[width=0.8\linewidth]{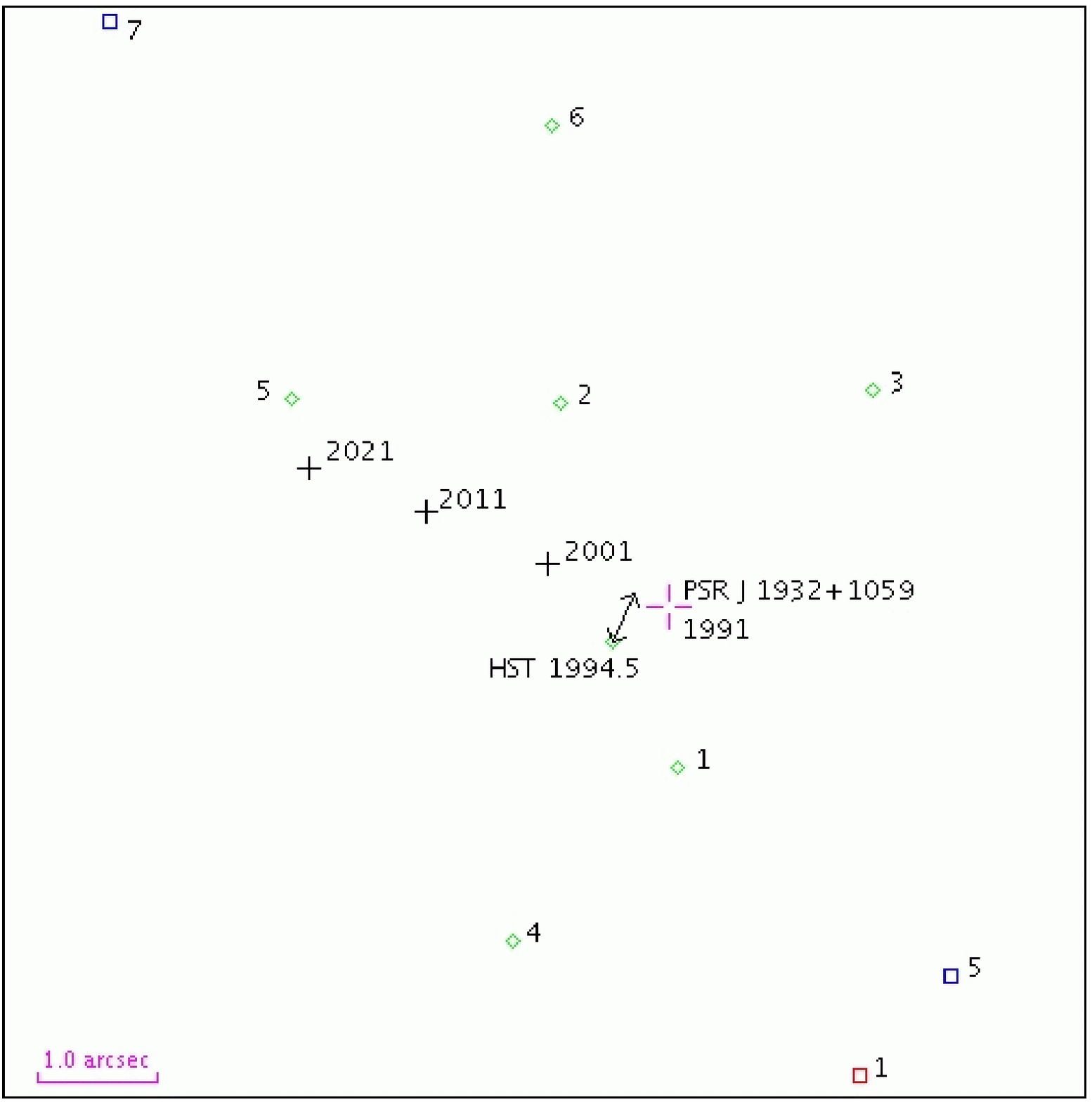}
\end{center}
\caption{\label{figps} 
Top panel: Objects from USNO-A2.0 (squares) and GSC2.2.1 (diamonds) within 
$9\arcsec$ from the 1991 position of PSR J1932+1059. We show also the estimated 
positions of PSR J1932+1059 for 2001, 2011 and 2021. 
Bottom panel: A magnification showing the positions of 6 stars and of the 
pulsar image (diamonds) as seen by the HST (Pavlov, Stringfellow \& C\'ordova 
\cite{Pavlov}). The USNO-A2.0 and GSC2.2.1 objects are now both shown as 
squares. The arrow shows the difference between the HST image and the 
estimated radio position at epoch 1994.5. 
The figures have been created with help of the Aladin Interactive Sky 
Atlas.} 
\end{figure}
%%%%%%%%%%%%%%%%%%%%%%%%%%%%%%%%%%%%%%%%%%%%%%%%%%%%%%%%%%%%%%%%%%%%%%%%%%%

Turning to the HST objects the numbers are much more encouraging.
Now the pulsar moves away from 1 towards 5 and stays at an almost constant 
distance to 2. Between 2010 and 2015 the image of 1 will move away from the
pulsar by $4 \mu$as, the image of 2 will move by $11 \mu$as almost 
parallel to the proper motion of the pulsar and the image of 5 will move 
by $12 \mu$as towards the pulsar. Thus the angular separation between 1 and 5 
should decrease by $8 \mu$as. If SIM is able to resolve the rather faint stars
1,2 and 5 a measurement of the astrometric lensing of a pulsar should be
possible. The best strategy would be to observe the field a few times per year
during the lifetime of SIM. 

In the above estimates the uncertainties from the distance estimate
are most crucial. We used a distance of $170$ pc based on the dispersion 
measure. This is not very reliable due to inhomogeneities within 
the interstellar medium. Several attempts to measure the parallax of
PSR J1932+1059 gave rise to a confusing situation. Salter et al.~(\cite{Sea}) 
obtained a parallax distance of $30 - 80$ pc, while Backer \& Sramek 
(\cite{BS}) found a lower limit of $250$pc. Allowing for the range 
$30 - 300$ pc implies an uncertainty of an order of magnitude for the 
predicted image shifts.   

\section{Conclusions}

We have argued that measuring the mass of stand-alone pulsars will 
be possible in the near future by means of gravitational microlensing.

The same strategy could also be used to measure the mass of isolated neutron 
stars, which are not seen as radio pulsars. Isolated neutron stars [for 
recent reviews see Treves et al.~(\cite{Treves}) and 
Popov et al.~(\cite{Popov})] 
may be detected as soft X-ray sources due to accretion in the case of old 
neutron stars or due to energy released in the cooling of young neutron stars.
Several candidates have been found by the 
ROSAT\footnote{\tt http://wave.xray.mpe.mpg.de/rosat} satellite. 
For two of those candidates, possible optical 
counterparts have been identified [\object{RX J185635-3754} by 
Walter \& Mathews (\cite{Walter}) and \object{RX J0720-3125} by 
Kulkarni \& van Kerkwijk (\cite{Kulkarni})], which makes them especially 
interesting. After the
submission of the first version of this work, Paczy\'nski (\cite{P01}) pointed
out that astrometric lensing of RX J185635-3754 might be possible 
with the HST in 2003.

In order to increase the list of possible candidates it is important to 
obtain as many proper motion and parallax measurements of pulsars and 
isolated neutron stars as possible. Candidates towards the galactic bulge 
should be especially considered with priority, since the chance of 
finding a large number of sources for lensing is maximised in that way.

We conclude that measuring neutron star (pulsar) masses by means of 
gravitational microlensing is possible. We suggest to observe 
\object{PSR J1932+1059} and three stars around it with SIM over several 
years to obtain such a mass measurement. 

{\it Note added in proof.} We learned after acceptance of this work that 
the possible mass determination of isolated pulsars by gravitational 
lensing has also been investigated in J.~E.~Horvath, 1996, MNRAS 278, L46.
Recently, a new parallax distance to PSR J1932+1059, 
$r = 331 \pm 10$ kpc, has been published by W.~F.~Brisken et 
al.\ {\tt astro-ph/0204105}; thus the astrometric effect is a factor
of 2 smaller than estimated in this work. Still, it should be possible 
to detect the astrometric lensing of PSR J1932+1059 with SIM.

\begin{acknowledgements}
We would like to thank W.~Kegel, D.~Lorimer, B.~Paczy\'nski and 
F.~Weber for very helpful conversations and references to the literature. 
This research has made use of the VizieR catalogue access tool and the 
Aladin Interactive Sky Atlas, CDS, Strasbourg, France. D.~J.~S.~would like 
to thank the Austrian Academy of Sciences for financial support.
\end{acknowledgements}

\end{document}